# Microlensing Events: End of the Dark Halo?

**Andrew Gould**

Dept of Astronomy, Ohio State University, Columbus, OH 43210

e-mail gould@payne.mps.ohio-state.edu

**Abstract**

I obtain an upper limit for the optical depth to microlensing toward Baade's Window of $3 \times 10^{-6}$ by assuming that all of the mass of the Galaxy interior to the Sun (and not in the bulge) is in a disk. The exponential scale height of the disk is left as a completely arbitrary function of radius and is varied to maximize the optical depth. I take account of the relatively small corrections induced by the fact that the bulge is not axisymmetric. If initial estimates by the OGLE collaboration of an observed optical depth $\tau \sim 3.3 \times 10^{-6}$ are confirmed, then essentially all of the dark matter interior to the Sun must be in a disk with a scale height of a few hundred parsecs.

Subject Headings: dark matter – gravitational lensing

# 1. Introduction

The OGLE collaboration (Udalski et al. 1994) has recently reported an optical depth to microlensing toward Baade's Window ($\ell = 1°$, $b = -4°$) of $\tau = (3.3 \pm 1.2) \times 10^{-6}$. This value is far higher than that expected from the known stars in the disk (Griest et al. 1991) and the bulge (Kiraga & Paczyński 1994). These two known components together account for only $\tau \sim 1.2 \times 10^{-6}$.

Here I show that even if all of the mass of the Galaxy interior to the Sun is placed in a disk, and even if the optical depth is maximized by varying the exponential scale height of the disk as a function of radius as a free parameter, the theoretically predicted optical depth toward Baade's Window only barely reaches the observed value. If the high optical depth found by OGLE is confirmed, then there is little if any room for a spherical dark halo.

The initial reports of observations by the MACHO collaboration (Alcock et al. 1994) also lead to a high optical depth. The MACHO group considers a wide range of double-exponential disk models and concludes that the high observed optical depths require an extremely flat dark matter distribution in accord with the suggestion originally made by Gould, Miralda-Escudé, & Bahcall (1994). The point of the present *Letter* is to demonstrate that this conclusion is a general result for axisymmetric disks and is not restricted to any particular class of models.

# 2. Absolute Upper Limit On Optical Depth

I assume that the disk of the Milky Way is circularly symmetric and that the mass at each radius $r$ is exponentially distributed in height $z$ above the plane with scale height $h(r)$. I define $\zeta(r)$ by

$$\zeta(r) \equiv \frac{1}{M_0} \frac{dM_{\text{disk}}(r)}{dr}, \qquad (2.1)$$

where $M_{\text{disk}}(r)$ is the total mass of the disk interior to $r$, and $M_0$ is the total mass of the galaxy interior to the solar galactocentric radius, $R_0$. Note that both $h$ and

$\zeta$ are completely arbitrary positive functions of $r$ constrained only by

$$\int_0^{R_0} dr\, \zeta(r) \leq 1. \tag{2.2}$$

The local density may therefore be written

$$\rho(r,z) = \frac{M_0}{4\pi r h(r)} \zeta(r) \exp\left[-\frac{|z|}{h(r)}\right]. \tag{2.3}$$

The optical depth to lensing can generically be evaluated as,

$$\tau = \int_0^{D_{\text{OS}}} d\,D_{\text{LS}}\, \frac{4\pi G \rho(D_{\text{LS}}) D_{\text{OL}} D_{\text{LS}}}{c^2 D_{\text{OS}}}, \tag{2.4}$$

where $D_{\text{OL}}$, $D_{\text{LS}}$, and $D_{\text{OS}}$ are the distances between the observer, lens, and source. For the case of lensing by disk objects of bulge stars seen at an angle $\alpha \ll 1\,\text{rad}$ above or below the plane, $D_{\text{OS}} = R_0$, $D_{\text{LS}} = r$, and $\alpha D_{\text{OL}} = |z|$. Hence, substituting equation (2.3) into equation (2.4), I find

$$\tau = \frac{GM_0}{\alpha R_0 c^2} \int_0^{R_0} dr\, \zeta(r) Q\left[\frac{\alpha(R_0 - r)}{h(r)}\right], \qquad Q(y) \equiv y e^{-y}. \tag{2.5}$$

The solar rotation speed $v = 220\,\text{km s}^{-1}$ is related to $M_0$ and $R_0$ by $v^2 = GM_0/R_0$. Equation (2.5) can therefore be written,

$$\tau = \frac{1}{\alpha} \frac{v^2}{c^2} \int_0^{R_0} dr\, \zeta(r) Q\left[\frac{\alpha(R_0 - r)}{h(r)}\right]. \tag{2.6}$$

Note that the function $Q$ obeys $Q(y) \leq e^{-1}$. Hence, equation (2.6) gives rise to the inequality,

$$\tau \leq \frac{1}{e\alpha} \frac{v^2}{c^2} \int_0^{R_0} dr\, \zeta(r) \leq \frac{1}{e\alpha} \frac{v^2}{c^2}, \tag{2.7}$$

where in the final step I have made use of equation (2.2).

## 3. Realistic Upper Limit

The first inequality in equation (2.7) is saturated only if the line of sight toward the bulge is always locally at one scale height. However, it is unrealistic to assume that the disk scale height rises linearly toward the galactic center. A more plausible assumption is that the scale height is constant with radius. If all the Galaxy mass is in the disk and the rotation curve is assumed flat, then $\zeta \sim 1/R_0$. I then find from equation (2.6) that

$$\tau = \frac{1}{\alpha} \frac{v^2}{c^2} w^{-1}[1 - (1+w)e^{-w}], \qquad w = \frac{\alpha R_0}{h}. \tag{3.1}$$

This expression reaches a maximum when $(1 + w + w^2)e^{-w} = 1$, i.e., $w \sim 1.79$. This implies

$$\tau \leq \frac{1}{3.35\alpha} \frac{v^2}{c^2}, \tag{3.2}$$

where equality holds if all of the galaxy mass is in the disk and if $h \sim 0.56\,\alpha R_0$. Note that the more realistic limit is lower than the absolute limit by a factor $\sim 0.81$.

## 4. Correction For Bulge Lensing

The optical depth to the bulge due to lensing by bulge stars depends slightly on the line of sight and the model of the bulge. However, for generic lines of sight and axisymmetric models where all the kinematically observed mass is in compact objects, the optical depth is typically $\tau_b \sim 6 \times 10^{-7}$. In particular, for a line of sight toward Baade's Window and a Kent (1992) model of the bulge, $\tau_b \sim 6.6 \times 10^{-7}$. The bulge is known to be elongated (Blitz & Spergel 1991; Weiland et al. 1994) with an estimated axis ratio $b/a \sim 0.75$ (Binney et al. 1991). If the long axis were pointing directly toward us, this would result in an increase in $\tau$ by a factor $\sim a/b$. Since the true orientation is probably closer to $\theta \sim 16°$ (Binney et al. 1991), I estimate the correction factor to be $\sim (a/b)\cos^2\theta + (b/a)\sin^2\theta \sim 1.3$. I therefore estimate the lensing by the bulge toward Baade's Window as $\tau_b \sim 1.6\,v^2/c^2$.

Of course, the bulge not only adds to the total optical depth, it also subtracts from the total amount of mass available to the disk. The Kent model of the bulge has a total mass $M_b = 0.17\,M_G$. Therefore, when one includes the bulge, the maximum optical depths due to the disk should be reduced by a factor $\sim 0.83$ relative to equations (2.7) and (3.2).

## 5. Comparison With The OGLE Results

The OGLE collaboration has measured the optical depth to Baade's Window $(b = -3°.9)$ of $\tau = (3.3 \pm 1.2) \times 10^{-6}$. According to equation (2.7) as corrected for bulge lensing (see § 4), the maximum optical depth toward Baade's Window is

$$\tau \leq 6.1 \frac{v^2}{c^2} \sim 3.3 \times 10^{-6}. \tag{5.1}$$

A more realistic limit obtained from equation (3.2) is

$$\tau \leq 5.2 \frac{v^2}{c^2} \sim 2.8 \times 10^{-6}. \tag{5.2}$$

Equation (5.2) saturates the limit only if *all* of the mass interior to the Sun is in the bulge and disk and only if the disk has a scale height $h \sim 300\,\mathrm{pc}$.

## 6. More General Disk Profiles

The disk models examined in § 2, are arbitrary in their radial profile and their scale height as a function of radius, but the form of their vertical distribution is fixed to be exponential. I now consider completely general profiles for the density

$$\rho(r,z) = \frac{M_0}{4\pi r h(r)} \zeta(r)\, F\left[\frac{|z|}{h(r)}\right], \tag{6.1}$$

where $\int_0^\infty dz F(z/h) = h$. Equation (2.6) remains valid provided that $Q$ is redefined as $Q(y) \equiv y F(y)$. For any physically viable disk profile, $Q$ will still have a maximum similar to the value $e^{-1}$ obtained for an exponential disk. For example, if 10% of the disk mass is in gas which lies near the plane (and does not lens) and the remainder of disk is in an isothermal population of stars, then $F(y) = \mathrm{sech}^2(y + \tanh^{-1} 0.1)$. In this case, $Q \leq 1.06\,e^{-1}$. Thus, use of the somewhat idealized exponential-disk profile does not significantly affect the calculation.

## 7. Conclusions

In order to explain the high optical depth reported by OGLE toward Baade's Window within the context of an axisymmetric Galaxy, it is necessary to assume that *all* of the mass interior to the Sun is in either the luminous bulge, the luminous disk, or a highly flattened dark structure with a scale height $h \sim 300\,\mathrm{pc}$. Corrections for the triaxiality of the bulge do not significantly affect this result. Thus, if the high optical depth reported by OGLE are confirmed, there is no room for a spherical dark halo within the inner 8 kpc of the Milky Way.

**Acknowledgements**: I would like to thank R. Blum and C. Han for useful and interesting discussions.